\documentclass[preprint2,twoside]{hwo}

\usepackage{graphicx}

\input{hwo.h}

\begin{document}

\title{\textbf{\LARGE The Heavy Element Enrichment History of the Universe}}
\author {\textbf{\large Eric Burns,$^{1}$ Jennifer Andrews$^2$}}
\affil{$^1$\small\it Department of Physics and Astronomy, Louisiana State University, Baton Rouge, LA 70803 USA}
\affil{$^2$\small\it Gemini Observatory/NSF's NOIRLab, 670 N. A'ohoku Place, Hilo, HI 96720, USA}

\author{\footnotesize{\bf Endorsed by:}
Robert	Szabo	(HUN-REN CSFK, Konkoly Observatory),
Brad	Cenko	(NASA/GSFC),
Paul	O'Brien	(University of Leicester),
Heloise	Stevance	(University of Oxford),
Ian	Roederer	(North Carolina State University),
Mark	Elowitz	(Network for Life Detection (NfoLD)),
Om Sharan	Salafia	(INAF - Osservatorio Astronomico di Brera),
Luca	Fossati	(Space Research Institute, Austrian Academy of Sciences),
Margarita	Karovska	(SAO),
Eunjeong	Lee	(EisKosmos (CROASAEN), Inc.),
Gijs	Nelemans	(Radboud University),
Igor	Andreoni	(University of North Carolina at Chapel Hill),
Filippo	D'Ammando	(INAF-IRA Bologna),
Pranav	Nalamwar	(University of Notre Dame),
Brendan	O'Connor	(Carnegie Mellon University),
Griffin	Hosseinzadeh	(University of California, San Diego),
Eliza	Neights	(George Washington University, NASA GSFC),
Endre	Takacs	(Clemson University),
Melinda	Soares-Furtado	(UW-Madison),
Maria Babiuc	Hamilton	(Marshall University),
Borja	Anguiano	(CEFCA),
Stéphane	Blondin	(Aix Marseille Univ/CNRS/LAM),
Frank	Soboczenski	(University of York \& King's College London),
Shivani	Shah	(North Carolina State University)
}



\begin{abstract}
    Understanding where elements were formed has been a key goal in astrophysics for nearly a century, with answers involving cosmology, stellar burning, and cosmic explosions. Since 1957, the origin of the heaviest elements (formed via the rapid neutron capture process; r-process) has remained a mystery, identified as a key question to answer this century by the US National Research Council. With the advent of gravitational wave astronomy and recent measurements by the James Webb Space Telescope we now know that neutron star mergers are a key site of heavy element nucleosynthesis. We must now understand the heavy element yield of these events as well as mapping when these mergers occurred back through cosmic time, currently thought to peak when the universe was half its current age. This requires an extremely sensitive ultraviolet, optical, and infrared telescope which can respond rapidly to external discoveries of neutron star mergers. We here describe how the Habitable Worlds Observatory can provide the first complete answer to one of the questions of the century. \textit{This article is an adaptation of a science case document developed for the Habitable Worlds Observatory.}    \\
    \\
\end{abstract}

%



\vspace{2cm}

\section{Science Goal}

The origin of the heaviest elements has been a mystery for nearly 60 years, before confirmation in 2017 that the mergers of neutron stars (which are the remnants of exploded massive stars) are a key contribution. These heavy elements are of interest for forming life like us as they include iodine, which was crucial for life as we know it on Earth. More generally, these elements may facilitate the development of complex life as the half-lives of these heavy elements allow for planetary cores to remain molten for billions of years, which can protect the planetary atmosphere from flares from its host star. Further effectively all complex life on Earth utilizes iodine in key biological processes. Since heavy element enrichment affects when Earth-like planets can form, this question is aligned with HWO's prime question.

The ingredients to produce the heaviest elements include i) highly dense, free neutrons, ii) an excess of neutrons to protons, and iii) specific cases of neutrino type luminosities (neutrinos can convert protons to neutrons or the reverse, with the wrong balance these can shift a plasma to violate item ii) \citep{burbidge1957synthesis,cameron1957origin}. Neutron stars are the densest matter in the universe, containing the mass of the Sun in the size of a city. Neutron stars contain tightly bound neutrons; however, when they are formed, destroyed, or significantly disrupted, the conditions for heavy element nucleosynthesis (more specifically, rapid neutron capture nucleosynthesis, generally referred to as ``r-process'' nucleosynthesis) can be met. Most cosmic explosions involved in neutron stars track the star formation rate history of the universe, which peaked early in the universe. However, thanks to the combined efforts of major US facilities including LIGO, Fermi, and James Webb, we know that neutron star mergers are a key source of heavy element nucleosynthesis. These events occur far later into the universe than other possible sources, so this may delay the onset of complex life until relativity recently.

\textit{Pathways to Discovery in Astronomy and Astrophysics for the 2020s}, the 2020 Astrophysics Decadal, emphasized \textit{New Messengers and New Physics} as a key science theme this decade \citep{national2021pathways}. Further, the corresponding priority area is \textit{News Windows on the Dynamic Universe} emphasizing time-domain and multimessenger astrophysics. This is probing the universe through time-critical observations across wavelengths and the inclusion of information from other probes of the cosmos, including gravitational waves, neutrinos, cosmic rays, and dust. While this is a broad topic, understanding neutron star mergers including their contribution to the heavy elements of the universe is clearly a priority source class in the Decadal.

Perhaps more relevant on the timescale of an HWO launch is \textit{Connecting Quarks with the Cosmos: Eleven Science Questions for the New Century} \citep{national2003connecting}. This document was created by the National Research Council, in response to the incoming Presidential administration's request, to understand the grandest questions in physics and astrophysics on a century timescale, in order to guide future US investments. One of the questions identified is \textit{How Were the Elements from Iron to Uranium Made?}. Of all eleven questions, this has perhaps seen the greatest progress, thanks to flagship NSF and NASA facilities, including the James Webb Space Telescope. To drive this point home, answering this question through the multidisciplinary study of neutron star mergers is highlighted in the long-range planning documents of multiple fields. In addition to the astrophysics Decadal, neutron star mergers are a key highlight in the future plans for particle physics \citep{2024arXiv240719176A}, gravitation \citep{evans2023cosmic}, nuclear science \citep{NSAC-LRP-2023}, with additional discussion in key documents for atomic science \citep{national2020manipulating} and plasma physics \citep{baalrud2020communityplanfusionenergy}. The study of these events is important to NASA, the NSF, and the DOE, and involves work on several flagship physics facilities \citep{burns2025multidisciplinary}.

The observations required to map the heavy element enrichment history of the universe due to neutron star mergers cannot be done with other current or planned facilities. Hubble does not have the necessary response time. Other planned ultraviolet telescopes cannot reach the necessary distances. Other planned ground telescopes will provide crucial information, but cannot provide the required ultraviolet information.

\section{Science Objective}
Neutron star mergers are currently detected deep into the universe by existing high energy monitors, and in the HWO era will be detected by the third generation International Gravitational Wave Network. The signature of heavy element nucleosynthesis is referred to as a kilonova, which is a quasithermal transient produced in the hours to weeks after merger. The first light to escape can be in the ultraviolet, when the plasma is hottest, before progressively cooling to a peak in infrared. Given the broadband nature of blackbody radiation, multi-epoch photometric and spectroscopic observations are needed across ultraviolet, optical, and infrared wavelengths. This approach is required to fully isolate the importance of possible heating sources, allowing for the isolation of the heat from radioactive decay, and ultimately a measure of the amount, velocity, and composition of heavy elements forged in each event. 

The hardest observation to make is the ultraviolet, which must also be performed as early as possible as signal in these wavelengths fade rapidly. Thus, rapid ($~$1-few hour) ultraviolet observations of well-localized neutron star mergers detected by the International Gravitational Wave Network third generation interferometers. This is required to understand source evolution of neutron star mergers, in particular the evolution of their heavy element (lanthanide, actinide) yield. For the majority of distant events detections will only be possible with photometric observations; however, spectroscopic measurements are inherently more powerful and are required for events close enough for high SNR detections. However, understanding of their particular usefulness will become more clear with the results from UVEX and advancements in the underlying theory. HWO being possibly capable of such observations would encourage the community to invest effort to understand what spectroscopic capabilities may bring us.

\begin{figure*}[ht!]
    \centering
    \includegraphics[width=1.0\textwidth]{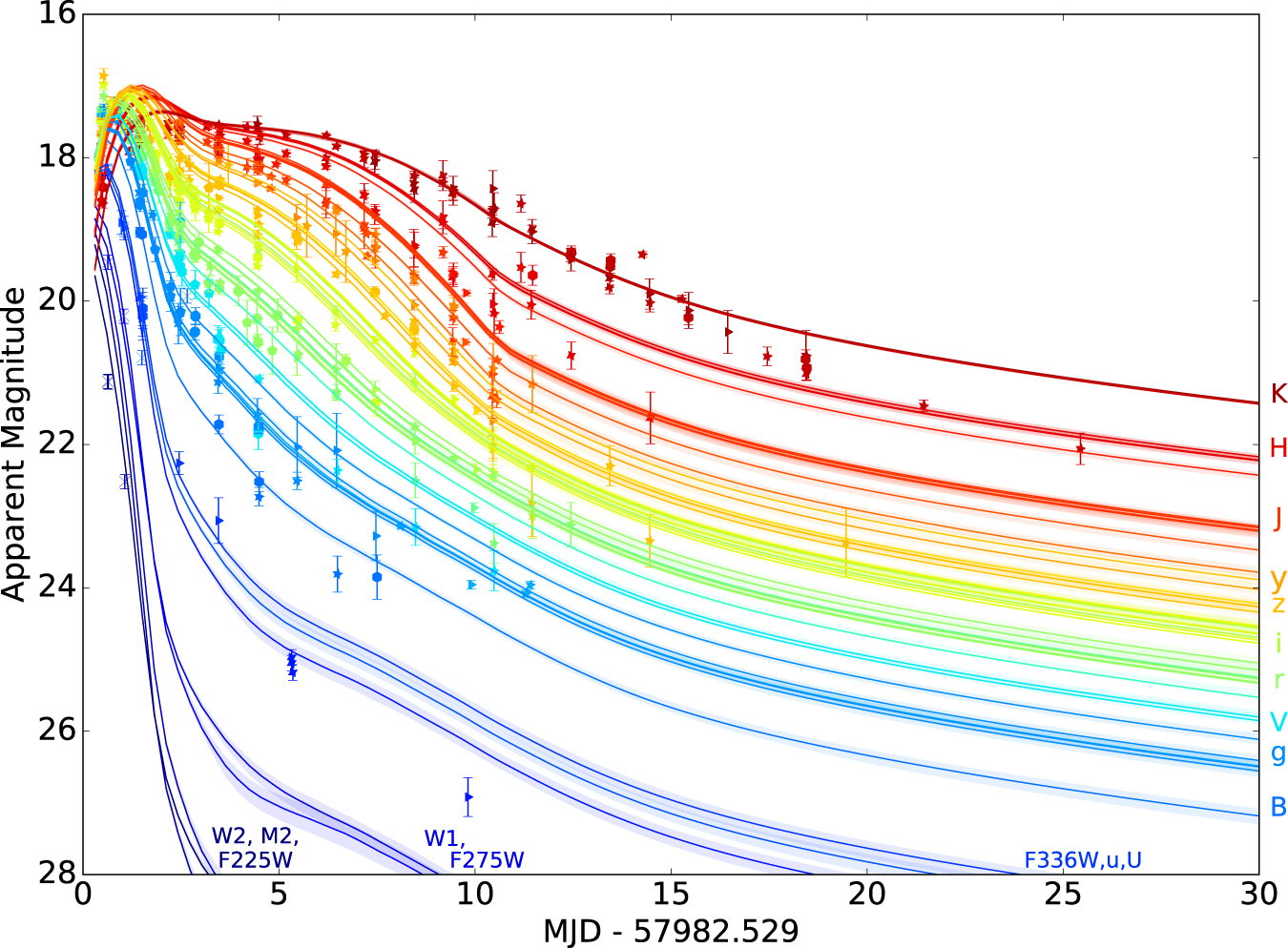}
    \caption{Image from \citealt{villar2017combined}, showing the composite ultraviolet, optical, and infrared observations of AT2017gfo, the kilonova found following the gravitation wave detection of a binary neutron star merger in 2017. The ultraviolet emission is brightest at early times and fades the most rapidly. While optical and infrared observations more directly track heavy element (r-process) yield, we must understand the physical origin of the ultraviolet emission in order to make proper inference on the physical properties of the ejecta (the amount of total mass, the elemental composition, the velocity distribution, etc).}
    \label{fig:villar}
\end{figure*}

\begin{figure*}[ht!]
    \centering
    \includegraphics[width=1.0\textwidth]{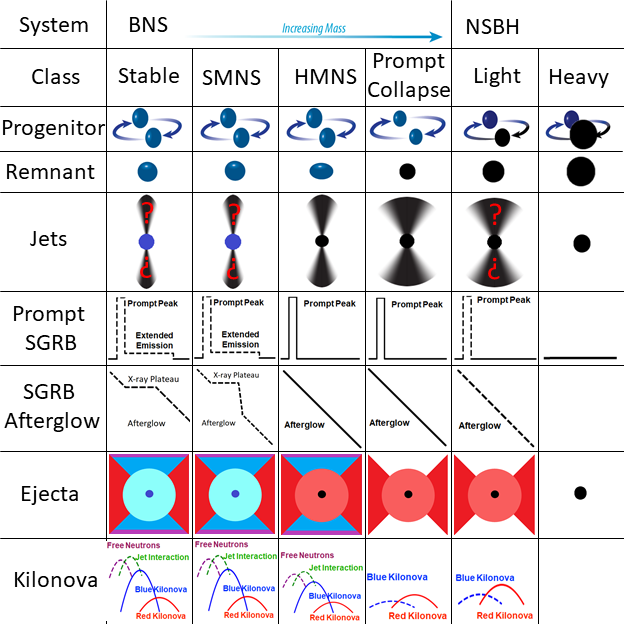}
    \caption{Figure 9 from \citealt{burns2020neutron}. The figure outlines the various observation signatures across the electromagnetic spectrum, which, when combined with gravitational wave measures of mass, can together disentangle the remnant objects immediately following a merger. This is deeply tied to the possible contributions to early ultraviolet emission, allowing for a way to determine the relative importance of these possible physical origins, and thus isolation of the contribution from the kilonova and ultimately measurement of the eject mass, velocity, and composition. HWO is required to map the relative frequency of the sources of ultraviolet signals in the low-mass cases, expected to occur most frequently.}
    \label{fig:lrr}
\end{figure*}

The early ultraviolet light from the first multimessenger detection of a binary neutron star merger in 2017, GW170817 / GRB 170817A / AT2017gfo, was a surprise. Its origin still eludes us. As summarized in \citet{burns2020neutron}, the possible contributions are varied. The ultraviolet, optical, and infrared transient is referred to as a kilonova. Its color is determined by numerous effects including composition, velocity distribution and asymmetries, ionization states of elements, and more. To highlight the effect of elemental composition, the bound-bound transitions of lanthanides and actinides blanket this energy range, causing a high opacity and generally precluding early light from escaping. If the early emission is from a low opacity kilonova then we can use these observations to constrain the total elemental yield released in the event.

Alternatively the early ultraviolet could be caused by the decay of free neutrons, by continued energy injection from the remnant object (e.g. a magnetar, or late-time fallback of weakly bound material onto the black hole engine), or interactions of the jet which powers the associated gamma-ray burst with the previously ejected material. Disentangling these requires early UV observations, see, e.g., \citet{metzger2018magnetar} and \citet{arcavi2018first}.

\begin{figure*}[ht!]
    \centering
    \includegraphics[width=1.0\textwidth]{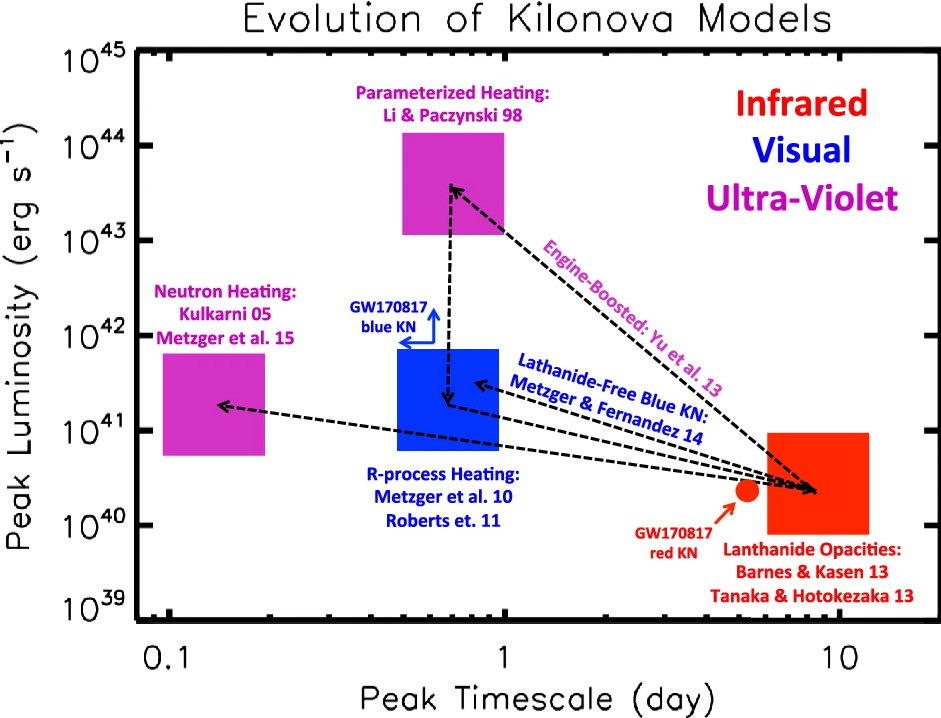}
    \caption{Figure 2 from \citet{metzger2020kilonovae}. While we know that early kilonova signals have significant ultraviolet emission, we do not know the physical cause. It may be possible to produce this signal purely from the radioactive decay of newly forged elements but this is difficult to explain with current models. Additional heating due to the decay of free neutrons or from the collisions of different ejected material (particularly the jet) can similar produce ultraviolet emission. However, these different origins produce different peak luminosities on different timescales, and can be distinguished with early ultraviolet observations.}
    \label{fig:lrr}
\end{figure*}

\section{Physical Parameters}
By observing the light output across the ultraviolet, optical, and infrared wavelengths and its evolution over time we can utilize kilonova models to infer the properties of the ejected material. Of key importance to our science goal is the amount of material ejected in a given explosion, as well as the composition of that ejecta, i.e., whether the heaviest elements comprise a low or high fraction.

There already exist advanced multiphysics models which tie observation to physical parameters. The capability and progress is evident in the Living Review on the topic \citep{metzger2017kilonovae} and the update three years late \citep{metzger2020kilonovae}. Since then, there have been continued meetings focusing on various aspects of the models, the input physics, and gains in knowledge from the detection of additional kilonovae. Thus, we have the required theory work, but we can also anticipate continued improvement until the launch of HWO. While spectra will likely be useful for more nearby events, the precise use of this data is not yet known, but will be in the coming years.

The existing literature focuses on photometric observations of nearby events, given the current state of UV instruments (Swift UVOT being the only rapid option) and the limited detection distance of the gravitational wave interferometers. The time of HWO will overlap with the next generation gravitational wave detectors, which can detect these events to a redshift of a few. Thus, extreme depth and reasonably rapid repointing is required. Additionally, spectra may provide additional discrimination possibilities, which should be explored in the literature.

The key thing to understand is the source evolution in terms of rates and in terms of elemental yield. The rates will be determined most directly by gravitational wave observations. The elemental yield requires characterization of the event, which requires HWO and similarly capable observatories (e.g. JWST, the ELTs), but the HWO ultraviolet is the most important. The average yield will likely change over cosmic time due to the different masses of the compact objects which merge, in non-trivial ways. Direct yield determination will always require electromagnetic characterization.

\begin{itemize}
    \item \textbf{State of the Art (SOA):} There are no early UV observations of neutron star mergers. This will change with the launch of UVEX in early 2030, which will provide key advanced insight into these events. However, these will be limited to the nearby universe (last two billion years), preventing understanding of how the UV emission of the population may change over cosmological timescales (i.e., source population evolution). HST will not perform follow-up faster than 48 hours.
    \item \textbf{Enhancing (incremental progress):} The first step would be detection of a small number of events at high redshift, to determine if there is a strong source evolution. The HWO observations can target events in specific redshift bins based on the information available in real-time from the gravitational wave alerts. The gravitational wave interferometers at the time of HWO will detect mergers at a rate of several per day, providing an easy sample to select from.
    \item \textbf{Enabling (substantial progress):} Basic source evolution requires measuring a small sample of events in a small number of redshift bins.
    \item \textbf{Breakthrough (major progress):} Moving out of low-count statistics ($~$10 events) in $~$8 bins would provide a well measured source evolution, giving us real understanding of one the heaviest elements first began to be formed, and when they began to approach the abundance necessary to enable life like us.
\end{itemize}

\begin{table*}[ht!]
    \centering
    \label{tab:obsreq}
    \begin{tabular}{lcccc}
        \noalign{\smallskip}
        \hline
        \noalign{\smallskip}
        {Observation} & {State of the Art} & {Incremental Progress} & {Substantial Progress} & {Major Progress} \\
        \hline
        Number of well-characterized objects & 0 & 5 & 25 & 80 \\
        \hline
    \end{tabular}\label{tab:req}
\end{table*}

\section{Description of Observations}
Detection of kilonovae at $<$30 mag over a range of photometric bands, emphasizing the ultraviolet bands, with observations beginning within hours and continuing every few hours until the ultraviolet has faded. The specific needs are summarized below in Figure~\ref{fig:hwo} and Table~\ref{tab:obsreq}

This is the key range to measure the onset of heavy element production based on current knowledge. Further, life first began on Earth about 0.8 billion years after formation, and humans only arose after 4.5 billion years after formation. The formation time is close to the peak merger rate. There is an additional delay of how long the heavy elements take to return to their galaxy, as mergers occur outside of their host galaxies, and is on the scale of a few billion years. Thus, while UVEX will give fantastic information on recent mergers, we want a full mapping of the cosmic history of mergers. Beyond a redshift of $~$1.5-2 the far ultraviolet emission will be redshifted to 400 nm or further, which is within range of the ground-based ELTs. Thus, HWO is critical in the range from $~$2-10 billion years ago, a timescale which includes the majority of mergers.

Rapid and multi-epoch spectroscopy for nearby mergers and multiband imaging for distant mergers would be transformational science. This drives the need for rapid commanding of HWO, with flexible scheduling. It drives an interest in on-board decision making for additional observations. E.g. if a given photometric band is brighter than a specific value then perform spectroscopic observations automatically, or if it is slightly less bright then take observations at adjacent bands. This will remove hours from the multi-epoch observing cycle by avoiding the need of down-link, processing, decision, then uplink.

\begin{figure*}[ht!]
    \centering
    \includegraphics[width=1.0\textwidth]{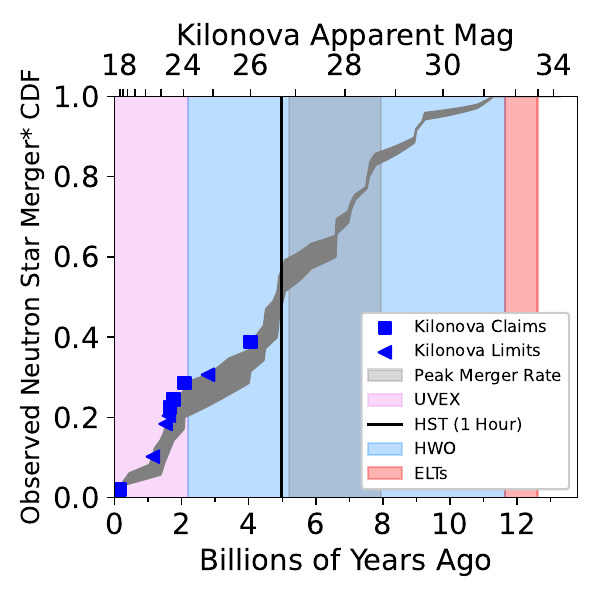}
    \caption{We have early imaging observations with Swift UVOT. There are no early spectroscopy observations. UVEX will bring both with incremental progress, launching around 2030. Observing with deep UV sensitivity in imaging, and early spectroscopy further into the universe, will be substantial progress. Spatial observations are 1” accuracy are sufficient. As the early observations will occur when the ejecta is moving at $>$0.1 c, where any lines would be broadened out, this source class will not drive spectral resolution requirements.}
    \label{fig:hwo}
\end{figure*}

\begin{table*}[ht!]
    \centering
    \label{tab:obsreq}
    \caption{Summary of the needed observations, current state of the art, and what is required for various levels of progress. Note that Swift UVOT enables on-target time on the order of minutes, but it is not sensitive enough for the science case described here.}
    \begin{tabular}{lcccc}
        \noalign{\smallskip}
        \hline
        \noalign{\smallskip}
        {Observation} & {State of the Art} & {Incremental Progress} & {Substantial Progress} & {Major Progress} \\
        \noalign{\smallskip}
        \hline
        \noalign{\smallskip}
        Imaging $<$2 days & Imaging 2-4 days & $<$26th mag in UV & Limiting $<$28th mag in UV & $<$30th mag in UV \\
         & Limiting mag $\sim$22 & 10 objects per epoch & 10 objects per epoch & 10 objects per epoch \\
        Spectroscopy $<$2 days & 2-4 days & 1 detection & 5 detection & 10 detections \\
        Rapid response &  2-4 days for HST \\
        \noalign{\smallskip}
        \hline
    \end{tabular}
\end{table*}


\bibliography{author.bib}

\end{document}